%% file: EAAMO23.tex
\documentclass{article}
\usepackage{arxiv}
\usepackage[textsize=tiny, textwidth=1.9in, colorinlistoftodos]{todonotes}

\usepackage{times}
\usepackage{soul}
\usepackage{url}
\usepackage{graphicx}

\usepackage[utf8]{inputenc}
\usepackage{amsmath}
\usepackage{amsfonts}  
\usepackage{amsthm}
\usepackage{booktabs}
\usepackage{algorithm}
\usepackage[switch]{lineno}
\usepackage{pgfplots} 
\usepackage{caption}
\usepackage{subcaption}
\usepackage{tikz}
\definecolor{vermillion}{RGB}{213,94,0}
\definecolor{SkyBlue}{RGB}{86,180,233}
\definecolor{BlueGreen}{RGB}{0,158,115}
\definecolor{orange}{RGB}{230,159,0}
\usetikzlibrary{positioning, matrix, fit, arrows}
\usepackage{adjustbox}
\usepackage{mathtools}
\usepackage{algpseudocode}
\usepackage{natbib}
\usepackage{soul}
\usepackage{url}
\usepackage[normalem]{ulem}
\usepackage[hidelinks]{hyperref}
\allowdisplaybreaks

\DeclareMathOperator*{\argminA}{arg\,min}

\newtheorem{theorem}{Theorem}[section]
\newtheorem{lemma}[theorem]{Lemma}

\title{Designing Equitable Transit Networks}

\author{
Sophie Pavia \\
Vanderbilt University \\
sophie.r.pavia@vanderbilt.edu
\And
J. Carlos Mart\'{i}nez Mori \\
Cornell University \\
jm2638@cornell.edu
\And
Philip Pugliese \\
CARTA \\
philippugliese@gocarta.org
\And
Samitha Samaranayake \\
Cornell University \\
samitha@cornell.edu
\And
Abhishek Dubey \\
Vanderbilt University \\
abhishek.dubey@vanderbilt.edu 
\And
Ayan Mukhopadhyay \thanks{Accepted in the non-archival track at the ACM Conference on Equity and Access in Algorithms, Mechanisms, and Optimization (EAAMO), 2023}\\
Vanderbilt University \\
ayan.mukhopadhyay@vanderbilt.edu 
}

\begin{document}

\maketitle

\begin{abstract}
    \input{abstract}
\end{abstract}

\input{introduction}

\input{model}
\input{experiments}

\input{conclusion}

\bibliographystyle{unsrtnat}
\bibliography{references}  

\clearpage
\pagenumbering{arabic}
\appendix

\end{document}

%% file: abstract.tex
Public transit is an essential infrastructure enabling access to employment, healthcare, education, and recreational facilities.
While transit accessibility is important in general, some segments of the population critically depend on it. 
However, existing paradigms for public transit design do not explicitly consider equity, which is often added as an additional objective \textit{post hoc}.
This design methodology hampers systemic progress toward equitable transit infrastructure. 
We present a mathematical formulation for transit network design that explicitly considers different notions of equity and welfare.
Our formulation is a mixed-integer linear program based on a (linearized) piece-wise linear utility function that quantifies the utility a passenger reaps from the installed network compared to the use of a personal vehicle.
We study the interaction between network design and different concepts of equity and showcase trade-offs based on real-world data from a metropolitan city in the United States of America. 

%% file: introduction.tex
\section{Introduction}
\label{sec: introduction}

Public transit enables people to access employment, healthcare, education, community resources, and recreation. 
An efficient public transit system helps members of communities interact with one another and aids the growth and expansion of businesses~\citep{federal2003status}. 
Public transit is not only important for individuals; it plays two essential roles in ensuring \textit{justice} to society: first, it distributes social and economic benefits, and second, it links the capabilities of the people, thereby enhancing what they can accomplish as a society~\citep{beyazit2011evaluating,harvey2010social}.

While access to transit infrastructure is vital in general, it is a more critical need for some people than others, i.e., a section of the population depends on public transit for their basic requirements (e.g., access to employment) more so than others~\citep{wp2015inequity}.
Indeed, in urban areas with low ridership, a higher percentage of passengers come from lower income levels due to fewer alternatives than those earning higher incomes~\citep{wp2015inequity}.
The importance of public transit for the lower-income population can be further understood from the glaring statistic that 90\% of the public assistance recipients do not own private vehicles and must rely on public transit for basic needs~\citep{federal2003status}.
As a result, it is imperative that policymakers design public transit infrastructure equitably by explicitly taking into account the diverse requirements (and dependencies) of people who use transit. 

In this paper, we investigate the problem of transit network design from the lens of equity.
We start with a clean slate, i.e., we assume that a transit designer has the scope to optimize a network from scratch. 
We point out that such optimization is typically infeasible in practice as most cities must optimize resources \textit{given} the current network. 
However, we believe that a fundamental analysis of transit design that explicitly focuses on equity and fairness is critical to shaping our future understanding of the intersection of network design and equity. 
We start with a background of related work on equity in transit and a brief overview of existing approaches to optimize transit network design.

\subsection{Background}
\label{sec: background}

\textbf{Different forms of equity in transit:} While it is well-understood that equitable transit is eminently desirable, it is not easy to precisely define what \textit{equity} means in this context. 
\citet{rock2014equity} as well as \citet{BEHBAHANI2019171} present a comprehensive overview of different philosophical trains of thought affecting the design of public transportation and point out that different notions of equity naturally lead to different outcomes. 
The major ethical theories relevant to public transportation are utilitarianism, egalitarianism, and sufficientarianism, each of which provides different guiding principles~\citep{van2011discussing}. 
Utilitarianism, broadly, focuses on maximizing the aggregate utility of a population~\citep{sen1979utilitarianism}. 
On the other hand, egalitarianism requires that all people are treated equally~\citep{rock2014equity}. 
The principle of sufficientarianism states that the distribution of a public good should be such that all recipients are sufficiently placed to meet their individual needs, even though the distribution might not be equal.

The concept of egalitarianism is particularly relevant to our context as it builds on the principle of social equality. There are several conceptualizations of egalitarianism. For example, the Rawlsian view of egalitarianism (or \textit{justice}, as \citet{rawls2004theory} referred to it in his seminal work) essentially posits that the distribution of public goods must improve the utility of the least advantaged population group~\citep{rawls2004theory}.
\citet{litman2017evaluating} explored how egalitarianism applies explicitly to public transit in one of the most comprehensive works on transportation equity and planning~\citep{litman2017evaluating} and proposed two forms of equity based on egalitarianism: horizontal equity, which focuses on equality of distribution irrespective of any attributes (e.g., income) of the individuals receiving the public good; and vertical equity, which is based on recognizing groups of people might have varying needs and favoring certain groups of people based on such attributes.
Several policy guidelines have been designed that recognize these notions of equity and the need for designing equitable transit systems by construction~\citep{litman2017evaluating,federal2003status,transitCenterReport}.

In this paper, we choose to explore the two most widely known and well-understood notions of equity in our analysis, i.e., the utilitarian view and the Rawlsian view. 
The model described in Section \ref{section: models for transit network design} can easily accommodate other notions of equity. 
We do not explore sufficientarianism in depth as it has been criticized in prior work in ethics~\citep{Paula2077ethics} and transportation~\citep{Martens2014IncorporatingEI}, since it is practically impossible to justify a strict sufficiency threshold.
This challenge can be overcome by using the concept of prioritarianism, i.e., using \textit{priority scores}, as in our approach.

\textbf{The current state of public transit in the United States (U.S.):} Despite the well-understood \textit{need} for equitable public transit~\citep{litman2017evaluating,beyazit2011evaluating,harvey2010social,rock2014equity, BEHBAHANI2019171}, principled approaches to \textit{design} equitable transit have received considerably lesser attention. 
We hypothesize that this gap is largely due to the lack of understanding of the interaction between transit network design and various notions of equity and fairness. As a result, existing public transit infrastructure often shows stark inequities. For example, 
a study on four major metropolitan areas in the U.S. found that low-wage workers have poorer access to employment opportunities via public transportation~\citep{stacy2020access}. 
Also, millions of elderly and people with disabilities have poor access to transit in the U.S.~\citep{degood2011aging}. 

Inequity can also exist in terms of the quality of transit. 
For example, in several counties in the U.S., affordable transit is built around poor infrastructure (longer travel times, no access to high-occupancy lanes, and narrow stops). 
In contrast, more expensive options are designed around comfort. 
As a result, while predominantly black passengers ride the former option, the majority of the passengers in the latter mode are white~\citep{spieler2020racism}. 
Large transit agencies in the U.S. are mandated by the Title VI of the Civil Rights Act of 1964~\citep{transitCenterReport}, which dictates that considerable change to existing infrastructure (e.g., fares and service frequency) must not have a disproportionate impact on people based on race, color, or national origin. 
However, these guidelines are narrow and lack robust instructions for implementation~\citep{transitCenterReport}.

\textbf{Transit network design:} The design of transit systems is a well-studied topic in transportation optimization.
However, because of the complexity of real-world transit operations, there is no single transit system design problem.
Instead, the design of transit systems is decomposed into a sequence of problems that consider incrementally operational considerations~\citep{desaulniers2007public,schobel2012line}.
For example, the first problem is to design the physical infrastructure network (e.g., where to install bus lanes), followed by the design of the operational network (i.e., line planning)~\citep{borndorfer2007column,schobel2012line}, to more nuanced problems such as frequency setting and pricing~\citep{bertsimas2020joint}, and finally crew and fleet scheduling~\citep{haase2001simultaneous}.
By and large, transit system design literature is concerned with a utilitarian guiding principle---the maximization of ridership.
While there are some exceptions to this view~\citep{camporeale2017quantifying,jiang2021reliability,rumpf2021public,CAMPOREALE2019184}, the efficiency and equity trade-offs of large-scale transit infrastructure networks remain insufficiently understood. 
Other approaches such as \citet{CAMPOREALE2019184} model equitable network design, however, their solution approach uses a metaheuristic, without any performance guarantees. 
We present a \textit{deterministic abstraction for network design with an exact solution method}.

\subsection{Contributions}
Our contributions are three-fold: \textbf{1)} We introduce an mixed-integer \emph{linear} programming (MILP) formulation for the design of transit networks that quantifies efficiency versus coverage trade-offs, accounting for the heterogeneous need for transit infrastructure. Our main technical ingredient is the design and linearization of a piece-wise linear \emph{utility function} that quantifies the utility reaped by passengers due to the installed network, as compared to the alternative: the use of personal vehicles. We emphasize that modeling this via an MILP (as opposed to a non-linear integer program) is particularly attractive from a scalability point of view. \textbf{2)} We evaluate network design using showcase our formulation by using real-world data from a city in the south of the U.S., namely Chattanooga. \textbf{3)} We present both qualitative and quantitative insights on the effects and trade-offs of using different notions of equity in network design. We expect that this study will serve as a starting point to make public transit design more equitable and fair for sections of society that need it the most.


%% file: model.tex
\section{Models for Transit Network Design}
\label{section: models for transit network design}

In this section, we introduce our models for transit network design.
Our models have identical solution spaces: a collection of binary decision variables indicating the parts of the network to be installed (subject to a budget constraint), together with real-valued variables quantifying the utility that residents who use public transit reap from the installed network.
However, they differ with respect to their objective functions, which reflect different models of social welfare.
We, therefore, refer to these objective functions as \emph{social welfare functions}: they measure the social welfare induced by a particular transit network design under a particular notion of welfare.

We focus on a simplified abstraction that captures the basic nature of network design\textemdash simultaneously connecting various pairs of nodes in a network\textemdash without the level of domain detail reserved for full-blown transit planning (e.g., capacities, frequencies, number of transfers).
This choice enables us to run comprehensive experiments shedding light on fundamental efficiency versus coverage trade-offs in a way that is more tractable and involves fewer model parameters. 
However, we note that our social welfare functions (our main modeling contribution) can be implemented similarly within more operational models of line planning, e.g.,~\citep{borndorfer2007column}).
As mentioned in Section~\ref{sec: background}, obtaining tractable models and algorithms for a transit network design is an active area of research in and of itself.

\subsection{Preliminaries}
\label{sec: preliminaries}
Let $G = (V,A)$ be a directed graph with lengths $\ell: A \rightarrow \mathbb{R}_{\geq 0}$ representing the underlying network on which a transit network may be installed (e.g., we may install dedicated bus lanes or other transit-related infrastructure on an existing road network).
Let $|V|$ and $|A|$ be denoted by $m$ and $n$, respectively.
There are costs $c: A \rightarrow \mathbb{R}_{\geq 0}$ associated with installing each arc as part of the transit network, as well as a budget $B \in \mathbb{R}_{\geq 0}$ which may not be exceeded.
We assume arcs are uncapacitated.
In this way, the solution space consists of all subsets $A_R \subseteq A$ forming a \emph{circulation}\footnote{
which is to say ``mass'' is conserved at every $v \in V$. 
This ensures transit routes are self-rebalancing.
} and for which $\sum_{a \in A_R} c_a \leq B$.

Let $\mathcal{D} = \{(o,d) \in V \times V: o \neq d\}$.
For each ordered pair $(o, d) \in \mathcal{D}$, which we refer to as an origin-destination pair, let $b_{od} \in \mathbb{N}_{\geq 0}$ represent the number of people who want to travel from $o$ to $d$ in a fixed time window (e.g., the morning rush hour).
We refer to the collection $\mathbf{b} = (b_{od})_{(o,d) \in \mathcal{D}} \in \mathbb{N}^{n(n-1)}$ as the travel demand profile. We assume that the demand profile is independent of the transit network design; our goal is to analyze how the design can best serve a given demand profile.
A social welfare function $w$ reflects the utility induced by a given transit network design under a particular notion of  welfare. In this case, social welfare reflects the aggregate (cumulative) level of service a transit network design offers, given the travel demand profile. 

We define level of service not as a binary quantity\textemdash \emph{given $A_R \subseteq A$ and an origin-destination pair $(o,d)$, is it possible to travel from $o$ to $d$ using only the arcs in $A_R$?}\textemdash but as a ``smoother'' quantity\textemdash \emph{given $A_R \subseteq A$ and an origin-destination pair $(o,d)$, how ``good'' is the travel from $o$ to $d$ using only the arcs in $A_R$?}
Here, we define ``good'' with respect to the level of service offered by an existing alternative: the use of personal vehicles. We choose to model the level of service in this manner as low-income households have lower rates of private car ownership, and mode choice against public transit is often driven by long travel times compared to using a private vehicle~\citep{o2018charting}.

Given an origin-destination pair $(o,d)$, let $\ell_{od}^* \in \overline{\mathbb{R}}_{\geq 0}$, where $\overline{\mathbb{R}} = \mathbb{R} \cup \{\infty\}$, denote shortest path distance (with respect to $\ell$) from $o$ to $d$ in $A$, where $\ell_{od}^* = \infty$ if and only if there is no path from $o$ to $d$ in $A$.
We assume $G$ is strongly connected, so in fact $\ell_{od}^* < \infty$ for all $(o,d)$.
Similarly, let $\ell_{od}(A_R) \in \overline{\mathbb{R}}_{\geq 0}$ denote shortest path distance from $o$ to $d$ in $A_R$, and note that, depending on $A_R$, it may very well be the case that $\ell_{od}(A_R) = \infty$.
For example, this trivially holds if $A_R = \emptyset$.
In any case, note that $\ell_{od}(A_R) \geq \ell_{od}^*$ for all $A_R \subseteq A$ and all $(o,d)$.

Let $\alpha \in \mathbb{R}_{\geq 1}$ be a model parameter reflecting the extent to which passengers tolerate detours; we define $\alpha$ as a multiplicative factor of the shortest path distance $\ell_{od}^*$.
Given $A_R$, we assume each passenger who wants to travel from $o$ to $d$ reaps unit \emph{utility} if $\ell_{od}(A_R) = \ell_{od}^*$, zero utility if $\ell_{od}(A_R) \geq \alpha \cdot \ell_{od}^*$, and otherwise reaps utility that interpolates linearly between the points $(\ell_{od}^*,1)$ and $(\alpha \cdot \ell_{od}^*,0)$. 
In this way, the passenger \emph{utility functions} are piece-wise linear functions that capture their level of service relative to their shortest path (which we assume they take if they use a personal vehicle).
More formally, for each $(o,d) \in \mathcal{D}$ we have $u_{od}: 2^A \rightarrow [0,1]$ where 
\begin{equation}
\label{eq: utility}
    u_{od}(A_R) \coloneqq
    \begin{cases}
    -\frac{\ell_{od}(A_R)}{\ell_{od}^* \cdot (\alpha - 1)} + \frac{\alpha}{(\alpha - 1)}, & \text{if } \ell_{od}^* \leq \ell_{od}(A_R) < \alpha \cdot \ell_{od}^*, \\
    0, & \text{otherwise}.
    \end{cases}
\end{equation}
Given $A_R \subseteq A$, let $\mathbf{u}(A_R) = (u_{od}(A_R))_{(o,d) \in \mathcal{D}} \in [0,1]^{n(n-1)}$ be the utility profile.
Then, a social welfare function $w: [0,1]^{n(n-1)} \rightarrow \mathbb{R}_{\geq 0}$ takes the utility profile $\mathbf{u}(A_R)$ (and implicitly the travel demand profile $\mathbf{b}$) to compute some aggregate measure of social welfare. For example, an aggregate measure of social welfare could simply be the cumulative welfare across all origin-destination pairs.

The remainder of this section is organized as follows.
In Section~\ref{sec: solution space} we present a mixed-integer linear programming model describing the solution space.
In Section~\ref{section: social welfare functions} we introduce two families of social welfare functions which, in addition to the utility profile $\mathbf{u}(A_R)$ and travel demand profile $\mathbf{b}$, also implicitly consider a priority profile $\mathbf{p} = (p_{od})_{(o,d) \in \mathcal{D}} \in (0,1]^{n(n-1)}$ that orders origin-destination pairs by level of service priority. For example, we may set $\mathbf{p}$ to favor network designs that best serve marginalized communities. 
An extension of the Rawlsian social welfare function is presented, as well, that assists in leftover resource allocation for budgets.
Lastly, in Section~\ref{section: model size reduction}, we describe a data-driven heuristic technique to reduce the size of our model.

\subsection{Solution Space}
\label{sec: solution space}

We extend a standard uncapacitated multi-commodity flow formulation~\citep{ahuja1988network} with additional decision variables and constraints that quantify the level of service, as given by \eqref{eq: utility}.
In particular, we have installation variables $x \in \{0,1\}^m$ indicating the installed network (referred to as $A_R \subseteq A$ in Section~\ref{sec: preliminaries}), connectivity variables $y \in \{0,1\}^{n(n-1)}$ indicating the origin-destination pairs to be connected, flow variables $f \in \{0,1\}^{m \cdot n(n-1)}$ delineating the paths through which said connectivity takes place (with $f^{od}_a$ denoting the flow on arc $a$ for the pair $(o,d) \in \mathcal{D}$), length variables $\ell \in \mathbb{R}_{\geq 0}^{n(n-1)}$ recovering the lengths of said paths (referred to as $(\ell_{od}(A_R))_{(o,d) \in \mathcal{D}}$ in Section~\ref{sec: preliminaries}), and utility variables $u \in \mathbb{R}_{\geq 0}^{n(n-1)}$ quantifying the level of service they offer (referred to as $\mathbf{u}(A_R) = (u_{od}(A_R))_{(o,d) \in \mathcal{D}}$ in Section~\ref{sec: preliminaries}). We maximize social welfare defined by the following set of constraints $P$:

\small
\begin{subequations}
\begin{align}
    && \sum\limits_{a \in A} c_a x_a &\leq B, & \label{constr: budget} \\
    && \sum\limits_{a \in \delta^+(i)} x_a &= \sum\limits_{a \in \delta^-(i)} x_a, & \forall i \in V \label{constr: mass} \\
    && \sum\limits_{a \in \delta^+(i)} f_a^{od} &- \sum\limits_{a \in \delta^-(i)} f_a^{od} = & \nonumber \\
    && \quad \quad y_{od} \cdot & \left(\mathbf{1}_{\{i = o\}} - \mathbf{1}_{\{i = d\}}\right), & \forall (o,d) \in \mathcal{D}, i \in V \label{constr: flow} \\
    && f_a^{od} &\leq x_a, & \forall (o,d) \in \mathcal{D}, a \in A \label{constr: installation} \\
    && \ell_{od} &= \sum\limits_{a \in A} \ell_a \cdot f_a^{od},  & \forall (o,d) \in \mathcal{D} \label{constr: length} \\
    && \ell_{od} &\leq (\alpha \cdot \ell_{od}^*) \cdot y_{od}, & \forall (o,d) \in \mathcal{D} \label{constr: detour} \\
    && u_{od} &= -\frac{\ell_{od}}{\ell_{od}^* \cdot (\alpha - 1)} \nonumber \\ 
    && & + \frac{\alpha}{(\alpha - 1)} \nonumber \\ 
    &&  & - \frac{\alpha}{\alpha-1} \cdot \left(1 - y_{od}\right), & \forall (o,d) \in \mathcal{D} \label{constr: utility} \\ 
    && x &\in \{0,1\}^m, & \nonumber \\
    && y &\in \{0,1\}^{n(n-1)}, & \nonumber \\ 
    && f &\in \{0,1\}^{m \times n(n-1)}, \nonumber \\
    && \ell &\in \mathbb{R}_{\geq 0}^{n(n-1)}, & \nonumber \\
    && u &\in [0,1]^{n(n-1)} & \nonumber
\end{align}
\end{subequations}
\normalsize
where, for any $i \in V$, let $\delta^+(i)$ and $\delta^-(i)$ denote the outgoing and incoming arcs of $i$ in $G$, respectively.
Constraint~\eqref{constr: budget} ensures the design remains within budget.
Constraints~\eqref{constr: mass} ensure mass conservation at every node.
Constraints~\eqref{constr: flow} are flow constraints ensuring a path from $o$ to $d$ whenever the origin-destination pair $(o,d) \in \mathcal{D}$ is served.
Constraints~\eqref{constr: installation} ensure flows can only use installed arcs.
Constraints~\eqref{constr: length} recover the lengths of the paths through which origin-destination pairs are connected.
Constraints~\eqref{constr: detour} ensure said lengths are ``tolerable'' (given the model parameter $\alpha \in \mathbb{R}_{\geq 1}$ described in Section~\ref{sec: preliminaries}).
Lastly, constraints~\eqref{constr: utility} implement the utility function from equation~\eqref{eq: utility}.

The reader might observe some differences between \eqref{eq: utility} and how it is implemented in $P$ (the set of constraints), and that the constraints $\ell_{od} \leq (\alpha \cdot \ell_{od}^*) \cdot y_{od}$ and $\ell_{od} = \sum_{a \in A} \ell_a \cdot f_a^{od}$ for all $(o,d)$ might lead to the infeasibility (under $P$) of solutions that are otherwise admissible under the abstract model described in Section~\ref{sec: preliminaries}.
This is by design, as we aim to model \eqref{eq: utility} with only linear constraints.
We now show that $P$ is a correct formulation.

\begin{lemma}
\label{lemma: model to ilp}
Let $A_R \subseteq A$ be any solution to the abstract model in Section~\ref{sec: preliminaries}.
Then, there exists a solution $(x, y, f, \ell, u) \in P$ with $u = \mathbf{u}(A_R)$.
\end{lemma}
\begin{proof}
Set $x_a = 1$ if and only if $a \in A_R$.
Clearly $\sum_{a \in \delta^+(i)} x_a- \sum_{a \in \delta^-(i)} x_a = 0$ for all $i \in V$ and $\sum_{a \in A} c_a x_a \leq B$.
Moreover, for each origin-destination pair $(o,d)$, set $y_{od} = 1$ if and only if $\ell_{od}(A_R) \leq \alpha \cdot \ell_{od}^*$.
If $y_{od} = 1$, set $\ell_{od} = \ell_{od}(A_R)$ and $f_a^{od} = 1$ if and only if arc $a \in A$ is in the shortest path from $o$ to $d$ in $A_R$ (breaking ties arbitrarily).
This assignment clearly satisfies all constraints involving $(o,d)$, and moreover leads to \eqref{eq: utility} coinciding (in the case where $\ell_{od}^* \leq \ell_{od}(A_R) < \alpha \cdot \ell_{od}^*$) with its implementation in $P$.
This shows that $u_{od} = u_{od}(A_R)$ whenever $y_{od} = 1$.
Conversely, if $y_{od} = 0$, set $u_{od} = \ell_{od} = 0$ and $f_{a}^{od} = 0$ for all $a \in A$.
This assignment clearly satisfies all constraints involving $(o,d)$.
Moreover, note that $u_{od}(A_R) = 0$ by \eqref{eq: utility}.
This shows that $u_{od} = u_{od}(A_R)$ whenever $y_{od} = 0$.
Therefore, $(x, y, f, \ell, u) \in P$ and $u = \mathbf{u}(A_R)$.
\end{proof}
The idea behind the proof of Lemma~\ref{lemma: model to ilp} is to offer no connectivity whatsoever (and hence offer zero utility) to the origin-destination pairs that, although connected by a solution $A_R \subseteq A$ to the abstract model in Section~\ref{sec: preliminaries}, find the level of service offered by said connectivity to be intolerable (and hence, by \eqref{eq: utility}, already have zero utility).
In this way, we avoid any infeasibility under $P$ while maintaining $u = \mathbf{u}(A_R)$. A partial converse to Lemma~\ref{lemma: model to ilp} is as follows.
\begin{lemma}
\label{lemma: ilp to model}
Let $w: [0,1]^{n(n-1)} \rightarrow \mathbb{R}_{\geq 0}$ be any monotonic increasing function\footnote{we say $w$ is monotonic increasing if, for any $x = (x_1, x_2, \ldots, x_n), y = (y_1, y_2, \ldots, y_n) \in [0,1]^n$ with $x_i \leq y_i$ for all $i \in [n]$ and at least one inequality strict, we have $w(x) < w(y)$.}.
Let $(x, y, f, \ell, u) \in P$ maximizing $w(u)$.
Then, there exists a solution $A_R \subseteq A$ to the abstract model in Section~\ref{sec: preliminaries} with $u = \mathbf{u}(A_R)$.
\end{lemma}
\begin{proof}
First, for any origin-destination pair $(o,d)$ with $y_{od} = 1$ and $\ell_{od} = \alpha \cdot \ell_{od}^*$, we set $y_{od} = 0$ and update all variables involving $(o,d)$ accordingly.
This preserves both feasibility and optimality (since $u$ does not change).
Now, set $A_R = \{a \in A: x_a = 1\}$.
Clearly $A_R \subseteq A$ forms a circulation and $\sum_{a \in A_R} c_a \leq B$.
Next, we claim that for any origin-destination pair $(o,d)$, $y_{od} = 0$ if and only if $\ell_{od}(A_R) \geq \alpha \cdot \ell_{od}^*$.
First, suppose $y_{od} = 0$ and assume by way of contradiction that $\ell_{od}(A_R) < \alpha \cdot \ell_{od}^*$.
Then, we can set $y_{od} = 1$, $f_a^{od} = 1$ for all arcs $a \in A_R$ along a shortest path from $o$ to $d$ in $A_R$ (this does not violate the constraints $f_a^{od} \leq x_a$ for all $(o,d)$ and $a$), and update $\ell_{od}$ and $u_{od}$ accordingly.
By our assumption that $\ell_{od}(A_R) < \alpha \cdot \ell_{od}^*$, this does not violate the constraint $\ell_{od} \leq (\alpha \cdot \ell_{od}^*) \cdot y_{od}$ and therefore preserves feasibility while yielding $u_{od} > 0$, contradicting optimality.
Conversely, suppose $\ell_{od}(A_R) \geq \alpha \cdot \ell_{od}^*$ and assume by way of contradiction that $y_{od} = 1$.
Then, the constraints $\ell_{od} = \sum_{a \in A} \ell_a \cdot f_a^{od}$ and $\ell_{od} \leq (\alpha \cdot \ell_{od}^*)$ are incompatible, contradicting feasibility.
Next, we claim that for any $(o,d)$ with $y_{od} = 1$, the edges $\{a \in A_R: f_{a}^{od} = 1\}$ form a shortest path from $o$ to $d$ in $A_R$.
If not, we can set $f_a^{od} = 1$ for all arcs $a \in A_R$ along a shortest path from $o$ to $d$ in $A_R$ (this does not violate the constraints $f_a^{od} \leq x_a$ for all $(o,d)$ and $a$) and $f_a^{od} = 0$ elsewhere, and update $\ell_{od}$ and $u_{od}$ accordingly.
This preserves feasibility while increasing $u_{od}$, contradicting optimality.
Therefore, $P$ implements \eqref{eq: utility} and $u = \mathbf{u}(A_R)$.
\end{proof}

The following is immediate.
\begin{theorem}
\label{theorem: correct}
Let $w: [0,1]^{n(n-1)} \rightarrow \mathbb{R}_{\geq 0}$ be any monotonic increasing function.
Then, $A_R \subseteq A$ is a solution to the abstract model in Section~\ref{sec: preliminaries} maximizing $w(\mathbf{u}(A_R))$ if and only if there exists $(x, y, f, \ell, u) \in P$ maximizing $w(u)$ with $u = \mathbf{u}(A_R)$. 
\end{theorem}
\begin{proof}
Follows from Lemma~\ref{lemma: model to ilp} and Lemma~\ref{lemma: ilp to model}.
\end{proof}

\subsection{Social Welfare Functions}
\label{section: social welfare functions}

We now describe three social welfare functions that aggregate level of service at the origin-destination level, as quantified by the utility profile $\mathbf{u}$ computed in Theorem~\ref{theorem: correct}, while implicilty accounting for the travel demand profile $\mathbf{b}$ and the priority profile $\mathbf{p}$. For any origin-destination pair $(o,d)$, we say the \emph{priority-adjusted} utility of a passenger traveling from $o$ to $d$ is $p_{od} \cdot u_{od} \in [0,1]$.
Then, if $b_{od} \in \mathbb{N}$ is the number of people who want to travel from $o$ to $d$, their total priority-adjusted utility is $b_{od} \cdot (p_{od} \cdot u_{od}) \in [0,b_{od}]$.
A priority-adjusted utilitarian social welfare function computes the sum of priority-adjusted utilities.
Therefore, the \emph{maximum priority-adjusted ridership} problem is 
\begin{equation}
\label{eq: max sum}
    \max_{(x, y, f, \ell, u) \in P} \sum_{(o,d) \in \mathcal{D}} b_{od} \cdot (p_{od} \cdot u_{od}).
\end{equation}
Note that the objective function in \eqref{eq: max sum} is monotonic increasing\footnote{we assume without loss of generality that $b_{od}, p_{od} > 0$ for all $(o,d) \in \mathcal{D}$.}, and so Theorem~\ref{theorem: correct} is applicable.
Similarly, the \emph{maximum priority-adjusted coverage} problem is
\begin{equation}
\label{eq: max min}
    \max_{(x, y, f, \ell, u) \in P} \min_{(o,d) \in \mathcal{D}} (1-p_{od}) \cdot u_{od}.
\end{equation}
The max-min nature of the formulation is based on the Rawlsian view of egalitarianism, i.e., we seek to maximize the utility of the least advantaged population group. While the objective function in \eqref{eq: max min} is not monotonic increasing as required by Theorem~\ref{theorem: correct}, we can make it so by including a small multiplicative factor of the objective function of \eqref{eq: max sum}.
In fact, this combination of the two welfare functions leads to a family of \emph{maximum priority-adjusted trade-off} problems, denoted by
\begin{subequations}\label{eq:combine}
    \begin{align}
        &\max_{(x, y, f, \ell, u) \in P} \Bigl\{ \gamma \cdot \sum_{(o,d) \in \mathcal{D}} b_{od} \cdot (p_{od} \cdot u_{od}) \quad + \nonumber\\
        &(1 - \gamma) \cdot \min_{(o,d) \in \mathcal{D}} (1-p_{od}) \cdot u_{od}\Bigl\} \tag{\ref{eq:combine}}
    \end{align}
\end{subequations}
where $\gamma \in (0,1]$.
In this way, the parameter $\gamma$ captures the priorities of the network planner by considering the trade-off between the two views of social welfare. 

It is important to observe that the max-min nature of \eqref{eq: max min} does not consider the welfare or utility of other groups once the maximin of the worst off is obtained. An extension of this, lexicographic maximization (leximax)~\citep{Chen2023}, alleviates the issue of leftover resources in the original max-min formulation by considering the welfare of all disadvantaged individuals. 
This process is done by iteratively solving a series of optimization problems for $k = 1, 2, \ldots, |\mathcal{D}|$. The process starts by setting $\mathcal{D}_0 = \mathcal{D}$. For notational ease, we refer to an arbitrary origin-destination pair as $(od)_i$, where $i$ is a quantifier. Now, at the $k$th iteration, let
\begin{equation}
\label{eq: leximax }
    (od)_k = \argminA_{(o,d) \in \mathcal{D}_k}  (1-p_{od}) \cdot u_{od}.
\end{equation}
That is, $(od)_k$ is the origin-destination pair that sets the ``floor'' in the $k$th iteration and is removed from the objective in subsequent iterations.
Formally,
\begin{equation}
    \mathcal{D}_k = \mathcal{D}_{k-1} \setminus \{(od)_{k-1}\},
\end{equation}
Here, $\mathcal{D}_k$ refers to the origin-destination pairs that remain in the objective function after the $k$th iteration.
We ensure the utility of the origin-destination pairs removed throughout the iterative process is conserved by collecting the value
\begin{equation}
    t_{(od)_k} = u_{(od)_k}
\end{equation}
and enforcing the constraint
\begin{equation}
    u_{(od)_k} \geq t_{(od)_k}
\end{equation}
in all subsequent iterations. Intuitively, we enforce that the OD pair that achieves the least utility in one iteration achieves at least the same utility in subsequent iterations and remove it from the objective function, thereby (potentially) enabling the optimization approach to find a higher lower bound (i.e., a higher floor). 
If there are two or more origin-destination pairs that achieve the minimum $\min_{(od) \in \mathcal{D}_k} (1-p_{od}) \cdot u_{od}$, we break ties by priority score, choosing the OD pair with the higher priority score. 
Thus, extending the Rawlsian formulation in order to maximize the welfare of the worst off, then the second worst off, and
so forth. 
This considers the welfare of
\textit{all} disadvantaged orgin-destination pairs by solving this sequence of optimization problems using our MILP.

\input{od}

%% file: od.tex
\subsection{Generating Origin-Destination Data and Model Size Reduction}
\label{section: model size reduction}

Although the description of $P$ involves polynomially many variables and constraints, it is still a very large formulation.
To see this, note that we require flow variables (denoted $\{f_a^{od}\}_{a \in A}$) for each origin-destination pair $(o,d)$.
Therefore, even if $m = \Theta(n)$, we require $\Theta(n^3)$ variables.
This is already computationally challenging for graphs with $n \approx 100$ (requiring in the order of a million variables), and becomes completely intractable for more realistic-sized graphs with $n \approx 1000$ (requiring in the order of a \emph{billion} variables).
Moreover, accounting for \emph{all} kinds of origin-destination pairs (as opposed to focusing on the ones with the highest travel demand) is necessary given the equity focus of our work.

To handle this difficulty, we propose the following conceptual approach to reduce the size of our model compared to explicitly modeling the entire road network. 
Recall that our goal is to generate data capturing a city's ground-truth travel pattern. One of the most critical needs of public transit is to enable access to employment. To capture travel patterns related to employment, we use the Longitudinal Employer-Household Dynamics Origin-Destination Employment Statistics (LODES ) data~\citep{lodes}, which is publicly available from the United States Census Bureau and captures the number of people traveling across census blocks for employment. We use this data to generate origin-destination data at the level of census tracts, i.e., our data is an aggregation of travel profiles across census tracts for employment (we describe the number of origin-destination pairs in section~\ref{section: experiments}).
To generate the underlying road network, we construct a graph whose vertices are denoted by the geographic centroid of each tract, and the edges represent the straight-line distance between the centroids.~\footnote{our framework can easily incorporate other notions of distance, e.g., computed through a mapping software.}

\subsection{Priority Scores}
\label{section: priority}

We use the notion of priority scores to capture the \textit{need} for transit, i.e., some sections of the community depend on transit more than others. We point out that the specific formulation of the score is outside the scope of our work; we leverage prior work~\citep{currie2004gap,currie2010quantifying} and use car ownership and household income as proxies for priority. First, we gather data pertaining to average  household income and for all census tracts from the American Community Survey Data (ACS)~\citep{acsIncome,acsCar}. We divide the spread of each attribute (e.g., income) into bins and assign a score (between 0 and 1) based on the percentile of the bin, i.e., the lowest bin is assigned a score of 0.1, and the highest bin is assigned a score of $1 - \epsilon$ for some small $\epsilon > 0$.
Then, for each census tract, we compute the sum of its car ownership score and income score. For example, assume that a tract falls in the second lowest bin concerning car ownership (i.e., a score of 0.2) and the lowest bin concerning income (i.e., a score of 0.1). The cumulative score for this census tract would be 0.3. Finally, we normalize the resulting scores across all tracts to create our proxy for priority. 
While this process gives us a priority score for each census tract, we still face two challenges.
First, recall that our model captures priority at the origin-destination level.
Second, analyzing the effect of network design on a large number of tracts (each with its own priority score) is cumbersome.
To tackle these challenges, we label each origin-destination pair with the priority score of the origin.

This assignment is based on the notion that we want to capture the need for transit at when residents travel \textit{to} the place of employment. 
Second, we create a set of priority classes by uniformly binning the range of priority scores; we refer to these partitions as priority groups. 
Each origin-destination pair, therefore, falls within one of these priority groups. 
It is important to note that \textit{our algorithm is agnostic to any binning of the origin-destination pairs into groups, as the optimization is performed at the level of OD pairs.} 
However, the resulting average utility within a group naturally depends on the grouping. 
Choosing the number of groups (or the methodology used for grouping) is outside the scope of our work, and we emphasize this must be done in partnership with policymakers, city planners, and social scientists.

%% file: experiments.tex
\begin{figure}[h]
\centering
\includegraphics[width=7cm]{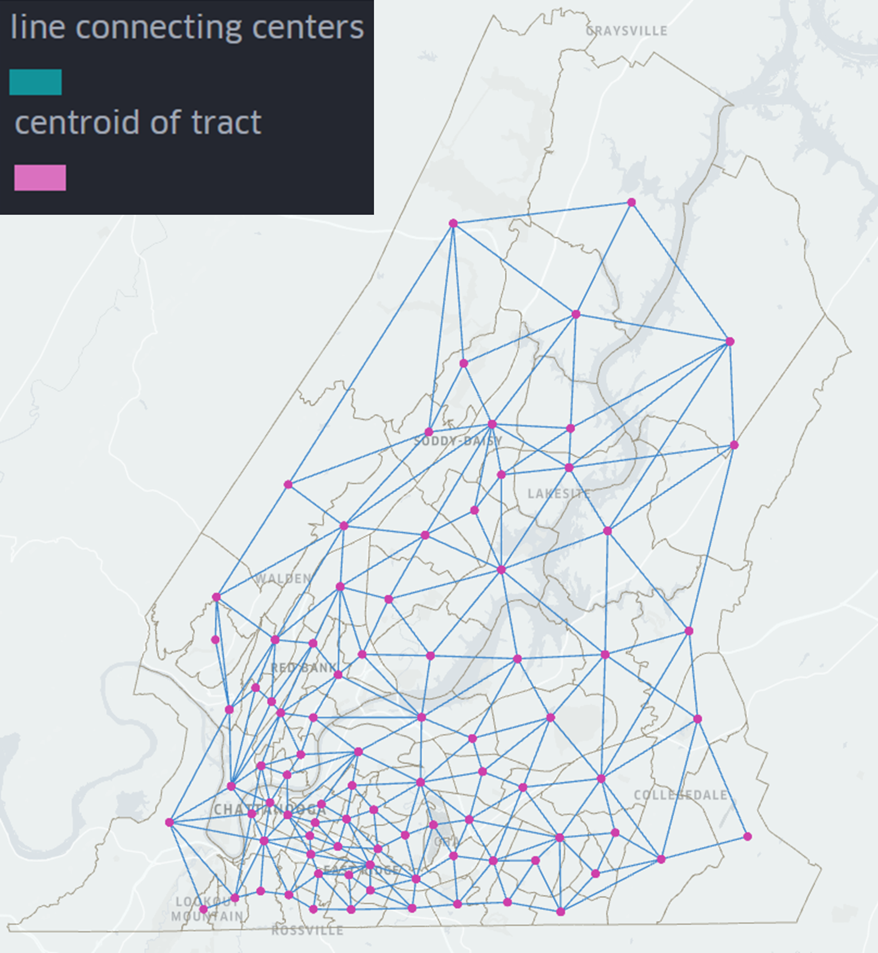}
\caption{Chattanooga, Tennessee, Hamilton County}
\label{fig:chatt}
\end{figure}

\section{Experiments}
\label{section: experiments}

\subsection{Experimental Setup}
Our implementation uses Python 3.10.6 and Gurobi 10.0.0~\citep{gurobi}. Our implementation and data are available as part of the supplementary material. For our experiments, we set $k=5$, i.e., we divide the range of priority scores into 5 distinct groups. Each origin-destination pair is therefore mapped to one of the groups. We also set $\alpha=2$, i.e., we set the passenger's detour tolerance to at most a factor of $2$.  We observe that the MILP solver can (at times) converge to solutions with high optimality gaps. To tackle this, as we iteratively increase the budget, we warm-start the solver with the solution from the last iteration. As the previous iteration uses a lower budget, its solution always remains feasible for subsequent iterations. Using the warm-start strategy, we generate experimental results to an optimality gap of $10^{-4}$.
All experiments were run on a Linux system having 112 Intel(R) Xeon(R) Gold 6238R CPU @ 2.20GHz and 512 GB of RAM.

\subsection{Area of Interest} The geographical area under consideration is Chattanooga, Tennessee, Hamilton County with a population of about 366,000. We consider a total of 46,008 origin-destination pairs for Hamilton county. 
We choose a smaller city as small cities have comparatively lower budgets available for transportation and incur significantly greater losses per trip compared to larger cities~\citep{kearney2015racing}. As a result, the need to provide equitable transit under strict budget considerations is critical. As we point out earlier, we perform experiments at the spatial granularity of census tracts; the proposed approach, however, could use any spatial discretization. Figure~\ref{fig:chatt} shows the geographical area of Chattanooga, Tennessee, Hamilton County divided into census tracts. The centers of census tracts are connected by the set of edges, $A$, and the installed network prescribes the set of edges to activate to serve the set of origin-destination pairs, $\mathcal{D}$, subject to the budget constraint. 

\subsection{Demand Distribution}
\input{plots/BarPlots.tex}

In Figure~\ref{fig:Demand}, we present the distribution of the travel demand among the priority groups in Chattanooga, Hamilton county. 
In our analysis, ``priority group 1'' is the highest priority group. 

\subsection{Results}
\input{plots/Chatt_main.tex}

\input{plots/BaseCase_Service.tex}
\input{results}

%% file: plots/BarPlots.tex
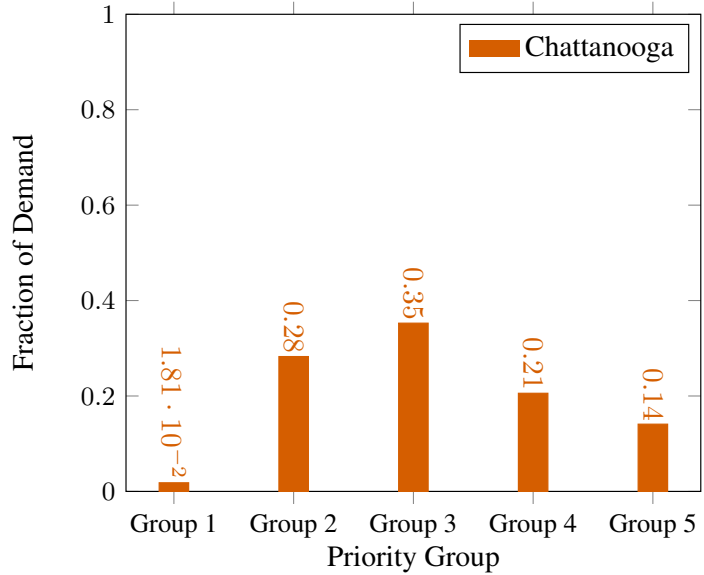
\begin{figure}[ht]
\centering
\resizebox{1\columnwidth}{!}{
\input{plots/line_plot/chattanooga/demand.tex}}
\caption{Distribution of Demand across Priority Groups}
\label{fig:Demand}
\end{figure}

%% file: plots/line_plot/chattanooga/demand.tex
\begin{adjustbox}{totalheight=0.075in}
\begin{tikzpicture}
\begin{axis}
[ 
ybar,
ymin = 0, ymax=1.0,
xtick = data,
  legend image code/.code={%
                    \draw[#1, draw=none] (0cm,-0.1cm) rectangle (0.6cm,0.1cm);
                },  
symbolic x coords={Group 1, Group 2, Group 3, Group 4, Group 5},
tick label style={font=\footnotesize}, 
ylabel=Fraction of Demand,
xlabel=Priority Group,
 nodes near coords,
    nodes near coords style={anchor=east,rotate=-90,inner xsep=1pt}
]
\addplot[ybar,fill, color=vermillion] table [x, y=d, col sep=comma] {csvs/demand_hist.csv};
\legend {Chattanooga};

\end{axis}
\end{tikzpicture}
\end{adjustbox}

 

%% file: plots/Chatt_main.tex
\begin{figure*}[h]
\centering
\begin{subfigure}[b]{.33\textwidth}
\resizebox{1\columnwidth}{!}{
\input{plots/line_plot/chattanooga/center/Utilitarian/EP.tex}}
\caption{Equal Priorities}
\end{subfigure}\hfill
\begin{subfigure}[b]{.33\textwidth}
\resizebox{1\columnwidth}{!}{
\input{plots/line_plot/chattanooga/center/Utilitarian/P.tex}}
\caption{Unequal Priorities}
\end{subfigure}\hfill
\begin{subfigure}[b]{.33\textwidth}
\resizebox{1\columnwidth}{!}{
\input{plots/line_plot/chattanooga/center/Utilitarian/difference.tex}}
\caption{Utility Gain} 
\end{subfigure}
\caption{Average Utility and Gain based on the Utilitarian Formulation: Chattanooga, Hamilton County, TN}
\label{fig:Utilitarian}
\end{figure*}
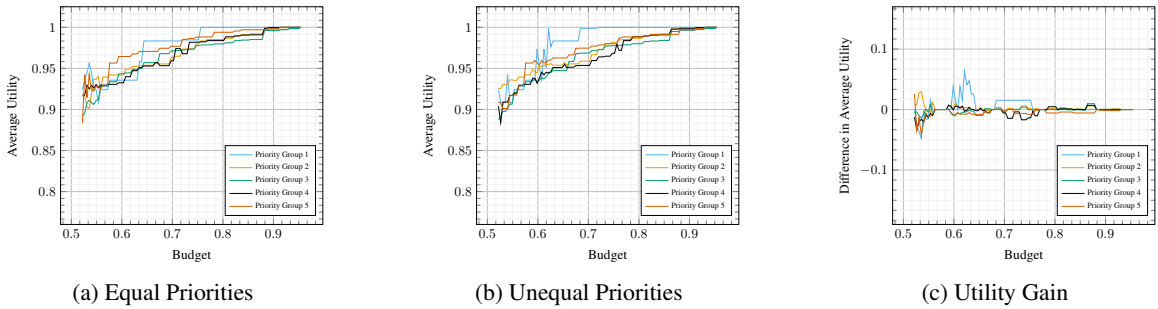

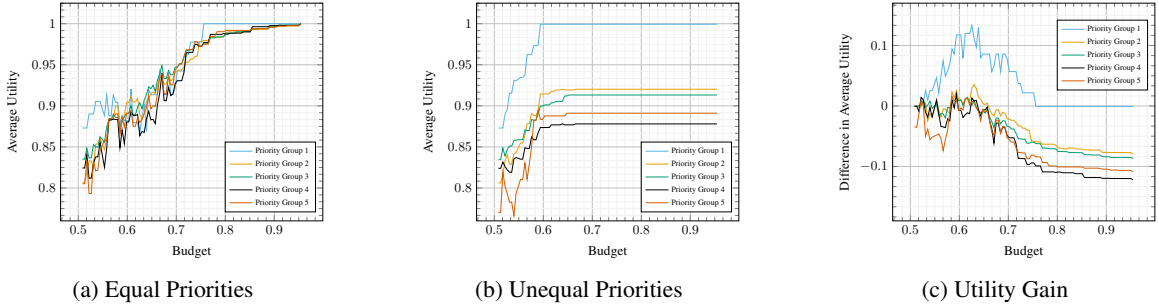
\begin{figure*}[h]
\centering
\begin{subfigure}[b]{.33\textwidth}
\resizebox{1\columnwidth}{!}{
\input{plots/line_plot/chattanooga/center/MaxMin/EP.tex}}
\caption{Equal Priorities}
\end{subfigure}\hfill
\begin{subfigure}[b]{.33\textwidth}
\resizebox{1\columnwidth}{!}{
\input{plots/line_plot/chattanooga/center/MaxMin/P.tex}}
\caption{Unequal Priorities}
\end{subfigure}\hfill
\begin{subfigure}[b]{.33\textwidth}
\resizebox{1\columnwidth}{!}{
\input{plots/line_plot/chattanooga/center/MaxMin/difference.tex}}
\caption{Utility Gain} 
\end{subfigure}
\caption{Average Utility and Gain based on the Rawlsian Formulation: Chattanooga, Hamilton County, TN}
\label{fig:MaxMin}
\end{figure*}

\begin{figure*}[h]
\centering
\begin{subfigure}[b]{.33\textwidth}
\resizebox{1\columnwidth}{!}{
\input{plots/line_plot/chattanooga/center/leximax/200_EP}}
\caption{Equal Priorities}
\end{subfigure}\hfill
\begin{subfigure}[b]{.33\textwidth}
\resizebox{1\columnwidth}{!}{
\input{plots/line_plot/chattanooga/center/leximax/200_EP_zoom}}
\caption{Equal Priorities: $k=1 \cdots 20$}
\label{zoom}
\end{subfigure}\hfill
\begin{subfigure}[b]{.33\textwidth}
\resizebox{1\columnwidth}{!}{
\input{plots/line_plot/chattanooga/center/leximax/200_P}}
\caption{Unequal Priorities}
\end{subfigure}
\caption{Results of Leximax Formulation when optimizing across ~80\% of the budget range: Chattanooga, Hamilton County, TN}
\label{fig:Leximax}
\end{figure*}
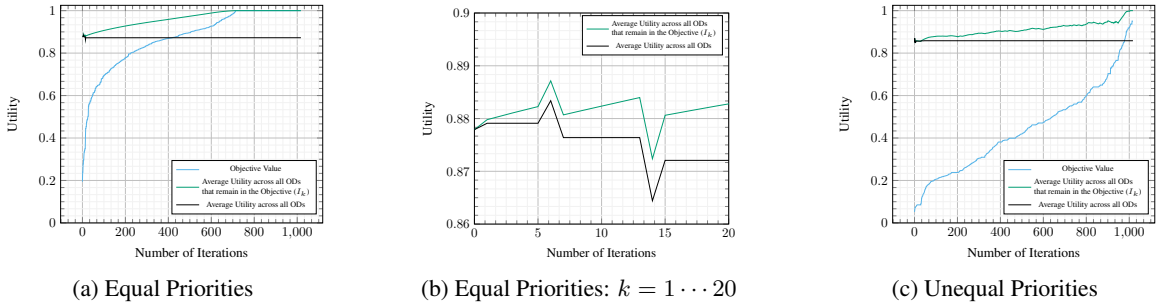

%% file: plots/line_plot/chattanooga/center/Utilitarian/EP.tex
\begin{adjustbox}{totalheight=0.2in}
\begin{tikzpicture}
\begin{axis}[
legend style={
            font=\tiny,
            at={(0.8,.04)},
            anchor=south,
        },
xlabel={Budget},
grid=both,
    grid style={line width=.1pt, draw=gray!10},
     minor tick num=5,
    major grid style={line width=.2pt,draw=gray!50},
ymin = 0.76, ymax = 1.025,
ylabel={Average Utility},
]
\addplot [color=SkyBlue] table [x=normal, y=p_1, col sep=comma] {csvs/chatt_center_U_EP.csv};
\addplot [color=orange] table [x=normal, y=p_2, col sep=comma] {csvs/chatt_center_U_EP.csv};
\addplot [color=BlueGreen] table [x=normal, y=p_3, col sep=comma] {csvs/chatt_center_U_EP.csv};
\addplot [color=black] table [x=normal, y=p_4, col sep=comma] {csvs/chatt_center_U_EP.csv};
\addplot [color=vermillion] table [x=normal, y=p_5, col sep=comma] {csvs/chatt_center_U_EP.csv};
\legend{Priority Group 1, Priority Group 2, Priority Group 3, Priority Group 4, Priority Group 5}
\end{axis}
\end{tikzpicture}
\end{adjustbox}

%% file: plots/line_plot/chattanooga/center/Utilitarian/P.tex
\begin{adjustbox}{totalheight=0.2in}
\begin{tikzpicture}
\begin{axis}[
legend style={
            font=\tiny,
            at={(0.8,.04)},
            anchor=south,
        },
xlabel={Budget},
grid=both,
    grid style={line width=.1pt, draw=gray!10},
     minor tick num=5,
    major grid style={line width=.2pt,draw=gray!50},
ymin = 0.76, ymax = 1.025,
ylabel={Average Utility},
]
\addplot [color=SkyBlue] table [x=normal, y=p_1, col sep=comma] {csvs/chatt_center_U_P.csv};
\addplot [color=orange] table [x=normal, y=p_2, col sep=comma] {csvs/chatt_center_U_P.csv};
\addplot [color=BlueGreen] table [x=normal, y=p_3, col sep=comma] {csvs/chatt_center_U_P.csv};
\addplot [color=black] table [x=normal, y=p_4, col sep=comma] {csvs/chatt_center_U_P.csv};
\addplot [color=vermillion] table [x=normal, y=p_5, col sep=comma] {csvs/chatt_center_U_P.csv};
\legend{Priority Group 1, Priority Group 2, Priority Group 3, Priority Group 4, Priority Group 5}
\end{axis}
\end{tikzpicture}
\end{adjustbox}

%% file: plots/line_plot/chattanooga/center/Utilitarian/difference.tex
\begin{adjustbox}{totalheight=0.2in}
\begin{tikzpicture}
\begin{axis}[
legend style={
            font=\tiny,
            at={(0.8,.04)},
            anchor=south,
        },
xlabel={Budget},
grid=both,
    grid style={line width=.1pt, draw=gray!10},
     minor tick num=5,
    major grid style={line width=.2pt,draw=gray!50},
ymin = -0.19, ymax = 0.17,
ylabel={Difference in Average Utility},
]
\addplot [color=SkyBlue] table [x=normal, y=p_1, col sep=comma] {csvs/difference_u_2.csv};

\addplot [color=orange] table [x=normal, y=p_2, col sep=comma] {csvs/difference_u_2.csv};

\addplot [color=BlueGreen] table [x=normal, y=p_3, col sep=comma] {csvs/difference_u_2.csv};

\addplot [color=black] table [x=normal, y=p_4, col sep=comma] {csvs/difference_u_2.csv};

\addplot [color=vermillion] table [x=normal, y=p_5, col sep=comma] {csvs/difference_u_2.csv};

\legend{Priority Group 1, Priority Group 2, Priority Group 3, Priority Group 4, Priority Group 5}
\end{axis}
\end{tikzpicture}
\end{adjustbox}

%% file: plots/line_plot/chattanooga/center/MaxMin/EP.tex
\begin{adjustbox}{totalheight=0.2in}
\begin{tikzpicture}
\begin{axis}[
legend style={
            font=\tiny,
            at={(0.8,.04)},
            anchor=south,
        },
xlabel={Budget},
grid=both,
    grid style={line width=.1pt, draw=gray!10},
     minor tick num=5,
    major grid style={line width=.2pt,draw=gray!50},
ymin = 0.76, ymax = 1.025,
ylabel={Average Utility},
]
\addplot [SkyBlue] table [x=normal, y=p_1, col sep=comma] {csvs/chatt_center_MaxMin_EP_ws.csv};

\addplot [color=orange] table [x=normal, y=p_2, col sep=comma] {csvs/chatt_center_MaxMin_EP_ws.csv};

\addplot [color=BlueGreen] table [x=normal, y=p_3, col sep=comma] {csvs/chatt_center_MaxMin_EP_ws.csv};

\addplot [color=black] table [x=normal, y=p_4, col sep=comma] {csvs/chatt_center_MaxMin_EP_ws.csv};

\addplot [color=vermillion] table [x=normal, y=p_5, col sep=comma] {csvs/chatt_center_MaxMin_EP_ws.csv};

\legend{Priority Group 1, Priority Group 2, Priority Group 3, Priority Group 4, Priority Group 5}
\end{axis}
\end{tikzpicture}
\end{adjustbox}

%% file: plots/line_plot/chattanooga/center/MaxMin/P.tex
\begin{adjustbox}{totalheight=0.2in}
\begin{tikzpicture}
\begin{axis}[
legend style={
            font=\tiny,
            at={(0.8,.04)},
            anchor=south,
        },
xlabel={Budget},
grid=both,
    grid style={line width=.1pt, draw=gray!10},
     minor tick num=5,
    major grid style={line width=.2pt,draw=gray!50},
ylabel={Average Utility},
ymin = 0.76, ymax = 1.025
]
\addplot [color=SkyBlue] table [x=normal, y=p_1, col sep=comma] {csvs/chatt_center_MaxMin_P_ws.csv};
\addplot [color=orange] table [x=normal, y=p_2, col sep=comma] {csvs/chatt_center_MaxMin_P_ws.csv};
\addplot [color=BlueGreen] table [x=normal, y=p_3, col sep=comma] {csvs/chatt_center_MaxMin_P_ws.csv};
\addplot [color=black] table [x=normal, y=p_4, col sep=comma] {csvs/chatt_center_MaxMin_P_ws.csv};
\addplot [color=vermillion] table [x=normal, y=p_5, col sep=comma] {csvs/chatt_center_MaxMin_P_ws.csv};
\legend{Priority Group 1, Priority Group 2, Priority Group 3, Priority Group 4, Priority Group 5}
\end{axis}
\end{tikzpicture}
\end{adjustbox}

%% file: plots/line_plot/chattanooga/center/MaxMin/difference.tex
\begin{adjustbox}{totalheight=0.2in}
\begin{tikzpicture}
\begin{axis}[
legend style={
            font=\tiny,
            at={(0.8,.6)},
            anchor=south,
        },
xlabel={Budget},
grid=both,
    grid style={line width=.1pt, draw=gray!10},
     minor tick num=5,
    major grid style={line width=.2pt,draw=gray!50},
ymin = -0.19, ymax = 0.17,
ylabel={Difference in Average Utility},
]
\addplot [color=SkyBlue] table [x=normal, y=p_1, col sep=comma] {csvs/chatt_center_MaxMin_diff_ws.csv};

\addplot [color=orange] table [x=normal, y=p_2, col sep=comma] {csvs/chatt_center_MaxMin_diff_ws.csv};

\addplot [color=BlueGreen] table [x=normal, y=p_3, col sep=comma] {csvs/chatt_center_MaxMin_diff_ws.csv};

\addplot [color=black] table [x=normal, y=p_4, col sep=comma] {csvs/chatt_center_MaxMin_diff_ws.csv};

\addplot [color=vermillion] table [x=normal, y=p_5, col sep=comma] {csvs/chatt_center_MaxMin_diff_ws.csv};

\legend{Priority Group 1, Priority Group 2, Priority Group 3, Priority Group 4, Priority Group 5}
\end{axis}
\end{tikzpicture}
\end{adjustbox}

%% file: plots/line_plot/chattanooga/center/leximax/200_EP.tex
\begin{adjustbox}{totalheight=0.2in}
\begin{tikzpicture}
\begin{axis}[
legend style={
            font=\tiny,
            at={(0.7,.04)},
            anchor=south,
            cells={align=left}
        },
xlabel={Number of Iterations},
grid=both,
    grid style={line width=.1pt, draw=gray!10},
     minor tick num=5,
    major grid style={line width=.2pt,draw=gray!50},
ymin = 0, ymax = 1.025,
ylabel={Utility},
]
\addplot [SkyBlue] table [x=output, y=obj, col sep=comma] {csvs/leximax_200_EP.csv};

\addplot [color=BlueGreen] table [x=output, y=avgU, col sep=comma] {csvs/leximax_200_EP.csv};

\addplot [color=black] table [x=output, y=avgU_a, col sep=comma] {csvs/leximax_200_EP.csv};

\legend{Objective Value, Average Utility across all ODs \\that remain in the Objective ($I_k$), Average Utility across all ODs}
\end{axis}
\end{tikzpicture}
\end{adjustbox}

%% file: plots/line_plot/chattanooga/center/leximax/200_EP_zoom.tex
\begin{adjustbox}{totalheight=0.2in}
\begin{tikzpicture}
\begin{axis}[
legend style={
            font=\tiny,
            at={(0.7,.8)},
            anchor=south,
            cells={align=left}
        },
xlabel={Number of Iterations},
grid=both,
    grid style={line width=.1pt, draw=gray!10},
     minor tick num=5,
    major grid style={line width=.2pt,draw=gray!50},
ymin = 0.86, ymax = 0.9,
xmin = 0, xmax = 20,
ylabel={Utility},
]
\addplot [color=BlueGreen] table [x=output, y=avgU, col sep=comma] {csvs/leximax_200_EP.csv};

\addplot [color=black] table [x=output, y=avgU_a, col sep=comma] {csvs/leximax_200_EP.csv};

\legend{Average Utility across all ODs \\ that remain in the Objective ($I_k$), Average Utility across all ODs}
\end{axis}
\end{tikzpicture}
\end{adjustbox}

%% file: plots/line_plot/chattanooga/center/leximax/200_P.tex
\begin{adjustbox}{totalheight=0.2in}
\begin{tikzpicture}
\begin{axis}[
legend style={
            font=\tiny,
            at={(0.7,.04)},
            anchor=south,
            cells={align=left}
        },
xlabel={Number of Iterations},
grid=both,
    grid style={line width=.1pt, draw=gray!10},
     minor tick num=5,
    major grid style={line width=.2pt,draw=gray!50},
ymin = 0, ymax = 1.025,
ylabel={Utility},
]
\addplot [SkyBlue] table [x=output, y=obj, col sep=comma] {csvs/leximax_200_P.csv};

\addplot [color=BlueGreen] table [x=output, y=avgU, col sep=comma] {csvs/leximax_200_P.csv};

\addplot [color=black] table [x=output, y=avgU_a, col sep=comma] {csvs/leximax_200_P.csv};

\legend{Objective Value, Average Utility across all ODs \\ that remain in the Objective ($I_k$), Average Utility across all ODs}
\end{axis}
\end{tikzpicture}
\end{adjustbox}

%% file: plots/BaseCase_Service.tex
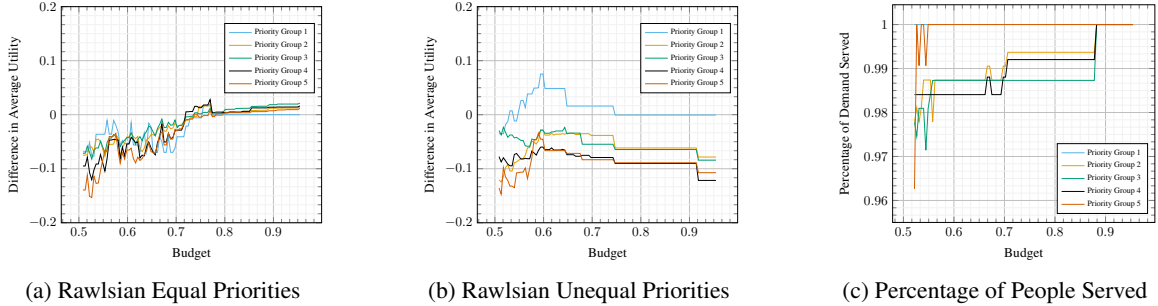
\begin{figure*}[ht]
\centering
\begin{subfigure}[b]{.33\textwidth}
\resizebox{1\columnwidth}{!}{
\input{plots/line_plot/chattanooga/center/difference/MaxMinEP_UtilEP.tex}}
\caption{Rawlsian Equal Priorities}
\label{r_ep}
\end{subfigure}\hfill
\begin{subfigure}[b]{.33\textwidth}
\resizebox{1\columnwidth}{!}{
\input{plots/line_plot/chattanooga/center/difference/MaxMinP_UtilEP.tex}}
\caption{Rawlsian Unequal Priorities}
\label{r_p}
\end{subfigure}\hfill
\begin{subfigure}[b]{.33\textwidth}
\resizebox{1\columnwidth}{!}{
\input{plots/line_plot/chattanooga/center/Utilitarian/service_EP.tex}}
\caption{Percentage of People Served} 
\label{u_ep}
\end{subfigure}
\caption{(\subref{r_ep}) (\subref{r_p})The gain in average utility compared to a utilitarian baseline, and (\subref{u_ep}) Percentage of people served by the utilitarian baseline in Chattanooga, Hamilton County, TN.}
\label{fig:ServeBase}
\end{figure*}


%% file: plots/line_plot/chattanooga/center/difference/MaxMinEP_UtilEP.tex
\begin{adjustbox}{totalheight=0.2in}
\begin{tikzpicture}
\begin{axis}[
legend style={
            font=\tiny,
            at={(0.8,.6)},
            anchor=south,
        },
xlabel={Budget},
grid=both,
    grid style={line width=.1pt, draw=gray!10},
     minor tick num=5,
    major grid style={line width=.2pt,draw=gray!50},
ymin = -0.2, ymax = 0.2,
ylabel={Difference in Average Utility},
]
\addplot [color=SkyBlue] table [x=normal, y=p_1, col sep=comma] {csvs/chatt_center_UMM_EP_diff.csv};

\addplot [color=orange] table [x=normal, y=p_2, col sep=comma] {csvs/chatt_center_UMM_EP_diff.csv};

\addplot [color=BlueGreen] table [x=normal, y=p_3, col sep=comma] {csvs/chatt_center_UMM_EP_diff.csv};

\addplot [color=black] table [x=normal, y=p_4, col sep=comma] {csvs/chatt_center_UMM_EP_diff.csv};

\addplot [color=vermillion] table [x=normal, y=p_5, col sep=comma] {csvs/chatt_center_UMM_EP_diff.csv};

\legend{Priority Group 1, Priority Group 2, Priority Group 3, Priority Group 4, Priority Group 5}
\end{axis}
\end{tikzpicture}
\end{adjustbox}

%% file: plots/line_plot/chattanooga/center/difference/MaxMinP_UtilEP.tex
\begin{adjustbox}{totalheight=0.2in}
\begin{tikzpicture}
\begin{axis}[
legend style={
            font=\tiny,
            at={(0.8,.6)},
            anchor=south,
        },
xlabel={Budget},
grid=both,
    grid style={line width=.1pt, draw=gray!10},
     minor tick num=5,
    major grid style={line width=.2pt,draw=gray!50},
ymin = -0.2, ymax = 0.2,
ylabel={Difference in Average Utility},
]
\addplot [color=SkyBlue] table [x=normal, y=p_1, col sep=comma] {csvs/chatt_center_UMM_P_diff.csv};

\addplot [color=orange] table [x=normal, y=p_2, col sep=comma] {csvs/chatt_center_UMM_P_diff.csv};

\addplot [color=BlueGreen] table [x=normal, y=p_3, col sep=comma] {csvs/chatt_center_UMM_P_diff.csv};

\addplot [color=black] table [x=normal, y=p_4, col sep=comma] {csvs/chatt_center_UMM_P_diff.csv};

\addplot [color=vermillion] table [x=normal, y=p_5, col sep=comma] {csvs/chatt_center_UMM_P_diff.csv};

\legend{Priority Group 1, Priority Group 2, Priority Group 3, Priority Group 4, Priority Group 5}
\end{axis}
\end{tikzpicture}
\end{adjustbox}

%% file: plots/line_plot/chattanooga/center/Utilitarian/service_EP.tex
\begin{adjustbox}{totalheight=0.2in}
\begin{tikzpicture}
\begin{axis}[
legend style={
            font=\tiny,
            at={(0.8,.04)},
            anchor=south,
        },
xlabel={Budget},
grid=both,
    grid style={line width=.1pt, draw=gray!10},
     minor tick num=5,
    major grid style={line width=.2pt,draw=gray!50},
ymin=0.955,
ylabel={Percentage of Demand Served},
]
\addplot [color=SkyBlue] table [x=normal, y=p_1, col sep=comma] {csvs/percentageServed_U_EP.csv};
\addplot [color=orange] table [x=normal, y=p_2, col sep=comma] {csvs/percentageServed_U_EP.csv};
\addplot [color=BlueGreen] table [x=normal, y=p_3, col sep=comma] {csvs/percentageServed_U_EP.csv};
\addplot [color=black] table [x=normal, y=p_4, col sep=comma] {csvs/percentageServed_U_EP.csv};
\addplot [color=vermillion] table [x=normal, y=p_5, col sep=comma] {csvs/percentageServed_U_EP.csv};
\legend{Priority Group 1, Priority Group 2, Priority Group 3, Priority Group 4, Priority Group 5}
\end{axis}
\end{tikzpicture}
\end{adjustbox}

%% file: results.tex
Note that the Rawlsian view of social welfare dictates that the max-min formulation is agnostic between serving less or more origin-destination pairs if the solution is unable to serve every pair; this is intuitive given the very nature of the max-min formulation. As a result, we plot the results starting from the least budget at which the max-min formulation can start serving everyone with some utility (the average utility in the max-min formulation is 0 for lower budgets). While the utilitarian formulation results in solutions with positive average utility for budgets less than this threshold, we plot results for the utilitarian formulation in the same budget range as the max-min formulation here in the main body of the paper for the sake of comparison. 
The highest budget that we use is the least budget with which all origin-destination pairs can be served through their shortest path, i.e., increasing the budget after this threshold makes no difference in the solutions. We plot our findings with respect to different fractions of this highest budget.

With respect to the leximax social welfare function, we perform experiments for select budgets that fall within the budget range used for the Rawlsian and Utilitarian experiments. The leximax social welfare function is run and plotted for iterations $k = 1, 2, \ldots, |\mathcal{D}|$, where $|\mathcal{D}|$ is the number of OD pairs in the network and the result presented is for a single budget.

\subsubsection{The Effect of Accounting for Priority}

We begin by considering the following question: \textit{how does the explicit consideration of priorities affect the average utility given a view of social welfare?}

\noindent \textbf{Utilitarian formulation:} We begin by comparing the average utility across the origin-destination pairs with and without the notion of priority (i.e., by treating all origin-destination pairs with the same priority), given the utilitarian objective. We present the results in Figure~\ref{fig:Utilitarian}. To visualize the difference made by accounting for priorities in our model, we plot the relative gain in average utility for each priority group when heterogeneous priority scores are considered compared to treating every origin-destination pair with the same priority. We observe that at lower budgets, most priority groups have a drop in average utility when priorities are explicitly modeled (notice most lines below 0 in the right-most plot of Figure~\ref{fig:Utilitarian} are at lower budgets). This loss occurs as the cumulative gain in utility (while not accounting for priorities) offsets any potential gain from considering the notion of priorities and maximizing the utility of groups with higher priority. Also, accounting for priority does not make a difference at higher budgets, as all groups can be served efficiently. However, with moderate budgets, we observe that accounting for priorities can significantly benefit groups with high priority.

\noindent \textbf{Rawlsian formulation:} We plot the results from the Rawlsian formulation in Figure~\ref{fig:MaxMin}. Similar to the utilitarian formulation, we observe that priorities do not make any difference at higher budgets, as all origin-destination pairs can be served efficiently. However, for lower budgets, explicitly accounting for utilities can provide significant gains to the priority group that needs public transit the most. It is important to note that the group that needs transit the most often gains at the expense of the other groups.

\noindent \textbf{Leximax formulation:} We present the results for the leximax formulation in Figure~\ref{fig:Leximax}. As we must vary the number of iterations for a fixed budget, we plot the results for 80\% of the total budget; our findings are, however, generalizable. Also, for ease of visualization, we plot the aggregated utility instead of the utility of each group; moreover,  our findings pertain more to the iterative optimization scheme of the leximax formulation than to the notion of priorities. We observe that iteratively solving the optimization problem can result in an increase in the resulting objective value. However, we point out two interesting findings. First, note that during successive iterations, the gain from ``raising the floor,'' i.e., improving the worst utility that any origin-destination pair can achieve, might come at the cost of a reduction in the average utility of all the ODs. We show this finding in Figure~\ref{fig:Leximax}\subref{zoom}. This observation is simply a consequence of the max-min formulation which only maximizes the minimum utility that an OD pair can achieve. However, we hypothesize that this finding has interesting policy implications, i.e., the trade-off between the decrease in average utility at the cost of improving the utility of the worst-served OD pair. Second, after a few iterations, the overall utility of the OD pairs does not change much, even though the utility of the OD pairs that remain as part of the objective value seems to continuously increase. At first glance, this finding is counter-intuitive; however, it is merely an artifact of the Leximax formulation. Note that the average utility of the OD pairs can increase even if the installed network does not change---as the worst OD pair (in terms of utility) is removed, the average utility of the remaining OD pairs goes up automatically.


\noindent \textbf{Key Takeaways:} Our observations are particularly relevant as we hypothesize that cities typically operate within moderate budget ranges, i.e., not too low for the quality of service to be poor but not high enough to provide everyone with public transit that is as good as using private vehicles. Our findings suggest that understanding which sections of a city \textit{need} public transit critically and accounting for such priorities during planning can improve accessibility for residents who depend on transit more than others. Moreover, using a Rawlsian view of welfare at low budgets can particularly benefit groups that need transit the most, keeping in mind their gain can be at the expense of other groups. The leximax formulation can be deployed after the Ralwsian formulation in order to remove the problem of leftover resources for a single budget. We emphasize that we \textit{do not prescribe} one conceptualization of equity over another. The challenge of choosing a notion of equity and computing priority scores are outside the scope of our work and must be determined by policymakers and social scientists.

\subsubsection{Comparing Against the Current Operational Paradigm}

Having observed the role of accounting for priority groups within each formulation, we now evaluate how different views of social welfare compare against the current operational paradigm of network design, i.e., maximizing ridership. We show the gain in utility with respect to the baseline for the Rawlsian formulation with and the without priorities.\footnote{Note that the comparison of the utilitarian formulation with the baseline is already shown in Figure~\ref{fig:Utilitarian}(c).} The question we seek to answer is: \textit{how do the proposed models perform relative to the ``business-as-usual'' model for maximizing ridership?}

We show the result in Figure~\ref{fig:ServeBase}. We observe that the Rawlsian formulation results in a loss of average utility for all priority groups when priorities are not considered (Figure~\ref{fig:ServeBase} (a)). At first glance, this finding looks counter-intuitive as it is against the very purpose of the Rawlsian formulation. However, this behavior is as expected because the Rawlsian formulation, by construction, results in a network that serves \textit{every} origin-destination pair with some utility. The baseline paradigm of maximizing ridership achieves higher cumulative utility by sacrificing service to some origin-destination pairs. We show the percentage of pairs served by the baseline in Figure~\ref{fig:ServeBase}(c). However, when priorities are considered, the regions of the city that need transit the most can gain with respect to the current operational paradigm (Figure~\ref{fig:ServeBase}(b)). 

\noindent \textbf{Key Takeaways:} Our findings suggest that while a Rawlsian view of social welfare results in serving every origin-destination pair, the solutions can result in lower cumulative utility with respect to the standard operational paradigm, especially when the notion of priorities is not considered. The average utility of the priority group that has a critical need for transit can improve by considering priorities. However, this gain comes at the cost of other priority groups. 
On the other hand, the utilitarian view (shown in Figure~\ref{fig:Utilitarian}(c)) can lead to a higher cumulative utility by sacrificing service to some regions of the city.

Policymakers can use our findings based on their domain expertise and requirements. For example, suppose there is a strict requirement to provide at least some level of public transit service (e.g., bus lines) to every region of the city. In that case, the Rawlsian formulation can be a natural choice for network design. On the other hand, if transit planners seek to maximize ridership using a utilitarian view of welfare, some additional form of service (e.g., through micro-transit or travel vouchers) can be provided to regions not served by the designed network directly. However, the policy implications are outside the scope of our work.

%% file: conclusion.tex
\section{Conclusion}

We present a mathematical formulation for transit network design that explicitly considers different notions of equity, welfare, and priority. 
Our formulation is a mixed-integer linear program based on a piece-wise linear utility function. Our experimental results demonstrate that considering the various degrees of \textit{needs} of the residents is critical to serving people who need transit the most. Our results also show that a utilitarian objective can achieve higher cumulative utility by sacrificing service to a small subset of origin-destination pairs. However, a Rawlsian view of welfare can be used to ensure that all regions are served (given at least some minimum budget), albeit at the cost of lower average utility. We hope our approach will help as a stepping stone for understanding the intersection of transit network design and social welfare considerations.